\documentclass[eqsecnum,noshowpacs,showkeys,nofootinbib,aps]{revtex4}
\usepackage{amsmath,amssymb}
\usepackage{epsfig}

\renewcommand{\theequation}{\arabic{equation}}
\newcommand\beq{\begin{equation}}
\newcommand\eeq{\end{equation}}
\newcommand\bea{\begin{eqnarray}}
\newcommand\eea{\end{eqnarray}}
\newcommand\nn{\nonumber}
\newcommand\pa{\partial}

\def\na{\nabla}

\begin{document}
\title{Baryon topology in hypersphere soliton model}
\author{Soon-Tae Hong}
\email{galaxy.mass@gmail.com}
\affiliation{Center for Quantum Spacetime and Department of Physics, Sogang University, Seoul 04107, Korea}
\date{\today}
\begin{abstract}
Exploiting a topological soliton on a hypersphere, we construct nucleon charge profile functions and find the density distributions for proton and neutron plotted versus the hypersphere third angle $\mu$. The neutron charge density 
is shown to possess a nontrivial $\mu$ dependence, consisting of both positive and negative charge density fractions. 
We next investigate the inner topology of the hypersphere soliton, by making use of the schematic M\"obius strips which are related with the tubular neighborhood of half-twist circle in the manifold $S^{3}$. In particular, we find that in the 
hypersphere soliton the nucleons are delineated in terms of a knot structure of two M\"obius strip type circles in $S^{3}$. 
Moreover, the two Hopf-linked M\"obius strip type circles in the hypersphere soliton are shown to correspond to (uu, d) in proton and 
(dd, u) in neutron, respectively, in the quark model.
\end{abstract}
\maketitle

\section{Introduction}
\setcounter{equation}{0}
\renewcommand{\theequation}{\arabic{section}.\arabic{equation}}

It is well known that the Dirac Hamiltonian scheme has been employed~\cite{di,pvn82,pvn89,hong15}, to convert the second class constraints into the 
first class ones. Moreover, making use of the Dirac quantization, there have been many 
attempts for quantizing the constrained systems, in order to obtain rigorous energy spectrum and 
investigate BRST~\cite{brst1,brst2,brst3} symmetries involved in the systems. 
In particular, the first order tetrad gravity has been investigated in the Dirac Hamiltonian scheme to see that 
the second class constraints reduce the theory to second order tetrad gravity and the first class constraints are different from those in the 
second order formalism, but satisfy the same gauge algebra if one uses Dirac brackets~\cite{pvn82}. There have been attempts to study the 
quantum BRST charge for quadratically nonlinear Lie algebras, and to derive a condition which is necessary and sufficient 
for the existence of a nilpotent BRST charge~\cite{pvn89}. In a (1+1) dimensional supersymmetric soliton, topological boundary conditions and the BPS bound have also been investigated to fix ambiguities in the quantum mass of the soliton~\cite{pvn99}. Moreover the Hamiltonian constraint approach has been applied to the cosmological perturbation theory~\cite{grav1,grav2}, and to the supersymmetric self-dual loop 
quantum cosmology~\cite{eder21}. 

On the other hand, it is well known that the standard Skyrmion model~\cite{skyrme61,anw83,hong98plb,hong15} is a constrained Hamiltonian system. In particular, the Skyrmion has been analyzed via the canonical quantization in the second class formalism, to study the baryon kinematics and its static properties~\cite{anw83}. Later, the hypersphere soliton model (HSM) has been proposed on 
the curved manifold~\cite{manton1} to obtain a topological lower bound on the baryon soliton energy and a set of equations 
of motion. Next we have newly predicted the baryon physical properties, such as masses, 
charge radii and magnetic moments in the HSM~\cite{hong98plb}.

Recently we have studied a mechanism of baryon mass genesis by exploiting a pulsating soliton in HSM~\cite{hong21}. To do this, first we have exploited the first class Hamiltonian to quantize the hypersphere soliton, and then have predicted  the axial coupling constant whose value is comparable to its corresponding experimental data. Next we have formulated the intrinsic frequencies of 
the pulsating baryons. Explicitly, we have evaluated the intrinsic frequencies $\omega_{N}$ and $\omega_{\Delta}$ of the 
nucleon and delta hyperon, respectively, to obtain the identity $\omega_{\Delta}=2\omega_{N}$.

In this paper we will consider the HSM, to study the baryon phenomenology and profile functions inside baryons. We will also construct 
nucleon charge profile functions and depict the density distributions for proton and neutron plotted versus the 
hypersphere third angle $\mu$. As a result, we will find that the neutron charge density has a nontrivial $\mu$ dependence.
Next, in the HSM, we will investigate the knot structure of the M\"obius strips associated with the tubular neighborhood of 
the half-twist circle inside the topological soliton. Moreover, we will consider two Hopf-linked~\cite{kauffman,baez} M\"obius strip type circles 
in the HSM, to show that they correspond to (uu, d) and (dd, u) combinations for the proton and neutron, respectively, in the quark model.

In Sec. II, we will briefly recapitulate baryon formalism to study the profile function inside the baryons. 
In Sec. III, we will analyze the charge profile functions of nucleons. 
In Sec. IV, we will investigate the knot structures of the M\"obius strip type half-twist circles inside the nucleons. 
Sec. V includes conclusions. In Appendix A, hypersphere geometry is summarized, 
and the Hopf fibration~\cite{hopf,manton2,bott} is discussed on the hypersphere.

\section{Baryon characteristics in HSM}
\setcounter{equation}{0}
\renewcommand{\theequation}{\arabic{section}.\arabic{equation}}

In this section, we exploit the HSM~\cite{manton1,hong98plb,hong21} to briefly study the baryon phenomenology and 
the profile function in Fig.~\ref{chargep}(a) in the identity map associated with the BPS bound in the baryon soliton energy.
To do this, we start with the Skyrmion Lagrangian density 
\beq
{\cal L}=\frac{f_{\pi}^{2}}{4}{\rm tr}(\partial_{\mu}U^{\dagger}
         \partial^{\mu}U)+\frac{1}{32e^{2}}{\rm tr}[U^{\dagger}\partial_{\mu}U,
         U^{\dagger}\partial_{\nu}U]^{2},
\label{lagtot}
\eeq
where $U$ is an SU(2) chiral field, and $f_{\pi}$ and $e$ are a pion decay constant and a dimensionless Skyrme parameter, respectively. 
Next the quartic term is necessary to stabilize the soliton in the baryon sector. 
Now we proceed to investigate baryon phenomenology by using the three-sphere metric given in Appendix A
\beq
ds^{2}=\lambda^{2}d\mu^{2}+\lambda^{2}\sin^{2}\mu~(d\theta^{2}+\sin^{2}\theta~d\phi^{2}).
\label{metrics30}
\eeq
Here the ranges of the hyperspherical coordinates are given by $0\le\mu\le\pi$, $0\le\theta\le\pi$ and $0\le\phi\le 2\pi$, and 
$\lambda$ ($0\le \lambda<\infty$) is the radius parameter of the hypersphere $S^{3}$. Note that $S^{3}$ is described in terms of the three dimensional 
hypersphere coordinates $(\mu,\theta,\phi)$ and the radius parameter $\lambda$.

In the hypersphere soliton on $S^{3}$, we obtain the soliton energy~\cite{manton1,hong98plb,hong21}
\beq
E=\frac{f_{\pi}}{e}\left[2\pi L\int_{0}^{\pi}d\mu\sin^{2}\mu\left(\left(\frac{d f}{d\mu}
    +\frac{1}{L}\frac{\sin^{2}f}{\sin^{2}\mu}\right)^{2}+2\left(\frac{1}{L}\frac{d f} 
     {d\mu}+1\right)^{2}\frac{\sin^{2}f}{\sin^{2}\mu}\right)
+6\pi^{2}\right],
\label{e2}
\eeq
where $L=ef_{\pi}\lambda$ ($0\le L<\infty$) is a radius parameter expressed in dimensionless units. Here $f(\mu)$ is the profile function for the hypersphere soliton, and satisfies $f(0)=\pi$ and $f(\pi)=0$ for unit baryon number. The above soliton energy has the BPS bound in the soliton model~\cite{manton1,hong98plb,hong21,faddeev,manton2} to yield the inequality
\beq
E\ge \frac{6\pi^{2}f_{\pi}}{e}|B|,\label{fbineq}
\eeq
where $B$ is given by~\cite{hong98plb}
\beq
B=-\frac{2}{\pi}\int_{0}^{\pi} d\mu \sin^{2} f \frac{df}{d\mu}.\label{bfmu}
\eeq
Note that $B$ is a conserved topological charge which is a baryon number in the HSM. 
For a single soliton, the topological charge is one~\cite{manton1,hong98plb,hong21,faddeev,manton2}.
In the cases of multi-Skyrmion solutions with $|B|>1$, the soliton energies will not satisfy the bound in Eq. (\ref{fbineq}) 
and possess much higher energy. The hedgehog solutions for the multi-Skyrmions then have been shown to be almost certainly all 
unstable saddle-points of the energy~\cite{manton2}.

The profile function $f$ in the soliton energy lower bound satisfies equations of motion~\cite{manton1,hong98plb},
\beq
\frac{d f}{d\mu}+\frac{1}{L}\frac{\sin^{2}f}{\sin^{2}\mu}=0,~~~\frac{1}{L}\frac{d f}{d\mu}+1=0.
\label{diff2}
\eeq
One of the simplest solutions of  (\ref{diff2}) is the identity map~\cite{manton1,hong98plb,hong21} 
\beq
f(\mu)=\pi-\mu\label{fmumap}
\eeq
with the condition $L=L_{B}$, where $L_{B}\equiv ef_{\pi}\lambda_{B}=1$. The profile function $f(\mu)$ in 
the identity map, associated with the BPS bound in the soliton energy, is depicted in terms of the angle $\mu$ 
in Fig.~\ref{chargep}(a). The fixed value $L=L_{B}$ can be used to 
obtain the soliton energy lower bound~\cite{manton1,hong98plb,hong21}\footnote{From now on we will use the condition $L_{B}=1$ to predict the physical quantities such as moment of inertia, baryon masses and charge radii in the HSM. We will then write the results for the identity map 
after the integral or definition expressions in (\ref{calipion}), (\ref{r2exp}) and (\ref{rhoexp}).} 
\beq
E=\frac{6\pi^{2}f_{\pi}}{e}.
\label{deltae}
\eeq

After performing the Dirac quantization in the first class formalism~\cite{di,hong15}, we obtain the modified 
Hamiltonian spectrum~\cite{hong15,hong21} 
\beq
\langle H\rangle=E+\frac{1}{2{\cal I}}\left[I(I+1)+\frac{1}{4}\right],
\label{canh}
\eeq
where $I$ $(=1/2,~3/2,...)$ are isospin quantum numbers of baryons. Here $E$ is given by  (\ref{deltae}) and ${\cal I}$ is 
the moment of inertia~\cite{hong98plb}
\beq
{\cal I}=\frac{8\pi}{3e^{3}f_{\pi}}\int_{0}^{\pi}d\mu\sin^{2}\mu\sin^{2}f
\left(1+\left(\frac{d f}{d\mu}\right)^{2}+\frac{\sin^{2}f}{\sin^{2}\mu}\right)=\frac{3\pi^{2}}{e^{3}f_{\pi}}.
\label{calipion}
\eeq
As a result, exploiting (\ref{canh}) we construct the Weyl ordering corrected nucleon and delta hyperon masses, respectively~\cite{hong21}
\beq
M_{N}=ef_{\pi}\left(\frac{6\pi^{2}}{e^{2}}+\frac{e^{2}}{6\pi^{2}}\right),~~~
M_{\Delta}=ef_{\pi}\left(\frac{6\pi^{2}}{e^{2}}+\frac{2e^{2}}{3\pi^{2}}\right)\label{mn2}.
\eeq
Next, in the HSM we obtain the electric isovector mean square charge radius $\langle r^{2}\rangle_{E,I=1}$ of the form~\cite{hong98plb}
\beq
\langle r^{2}\rangle_{E,I=1}=\frac{8\pi}{3 e^{2}f_{\pi}^{2}{\tilde{\cal I}}}\int_{0}^{\pi}d\mu\sin^{4}\mu\sin^{2}f
\left(1+\left(\frac{d f}{d\mu}\right)^{2}+\frac{\sin^{2}f}{\sin^{2}\mu}\right)=\frac{5}{6e^{2}f_{\pi}^{2}},
\label{r2exp}
\eeq
where $\tilde{{\cal I}}$ is the dimensionless moment of inertia defined as $\tilde{{\cal I}}=e^{3}f_{\pi}{\cal I}$. 
Similarly, we find the other charge radii in the HSM to yield~\cite{hong98plb,hong21}
\beq
\langle r^{2}\rangle^{1/2}_{E,I=1}
=\langle r^{2}\rangle^{1/2}_{M,I=0}=\langle r^{2}\rangle^{1/2}_{M,I=1}
=\langle r^{2}\rangle^{1/2}_{M,p}=\langle r^{2}\rangle^{1/2}_{M,n}
=\sqrt{\frac{5}{6}}\frac{1}{ef_{\pi}},\label{radii4}\\
\eeq
where the subscripts $E$ and $M$ stand for the electric and magnetic radii, respectively.

Making use of the charge radii in  (\ref{radii4}) we fix the value of $\langle r^{2}\rangle_{M,p}^{1/2}$ with the corresponding 
experimental data $\langle r^{2}\rangle_{M,p}^{1/2}=0.80$ fm. We then have 
\beq
ef_{\pi}=225.23~{\rm MeV}=(0.876~{\rm fm})^{-1}.\label{efpi}
\eeq 
Note that, inserting $ef_{\pi}=(0.876~{\rm fm})^{-1}$ into the condition $L_{B}=1$, 
we obtain $\lambda_{B}=0.876~{\rm fm}$. The hypersphere soliton is thus defined 
in terms of  the fixed radius parameter $\lambda_{B}$ which is comparable to the predictions for the charge radii 
in  (\ref{radii4}). In other words, in the HSM the universe ${\cal U}$ is given by a sum of two subspaces, namely 
${\cal U}=V_{B}(\lambda\le\lambda_{B})\oplus\bar{V}_{B}(\lambda>\lambda_{B})$ where $V_{B}(\lambda\le\lambda_{B})$ and 
$\bar{V}_{B}(\lambda>\lambda_{B})$ are the volume subspaces inside and outside the hypersphere soliton, respectively. 
Moreover, in HSM the baryons are described in terms of $V_{B}$ which is a three dimensional volume defined inside 
$S^{3}$ having the volume element in (\ref{volumnel}). Note that most of baryon charges are located inside the 
volume $V_{B}$ of radius parameter $\lambda_{B}$ as shown in  (\ref{radii4}). Explicitly 
$\langle r^{2}\rangle^{1/2}_{M,p}=\sqrt{\frac{5}{6}}\lambda_{B}=0.91~\lambda_{B}$, for instance.

Finally note also that the hypersphere coordinates $(\mu,\theta,\phi)$ are integrated out in  (\ref{e2}), 
and $E$ in  (\ref{deltae}) is a function of $\lambda_{B}=\frac{1}{ef_{\pi}}$ or equivalently $f_{\pi}$ and $e$ only. 
Moreover, similar to $E$ in (\ref{e2}) and $B$ in (\ref{bfmu}), the above physical quantities such as the moment of 
inertia and charge radii are also given by integral expressions, as in (\ref{calipion}) and (\ref{r2exp}) for instance. 
These physical quantities are thus obtained in terms of $f_{\pi}$ and $e$ only.

\section{Charge density profiles in HSM}
\setcounter{equation}{0}
\renewcommand{\theequation}{\arabic{section}.\arabic{equation}}

\begin{figure}[t]
\centering
\includegraphics[width=5.2cm]{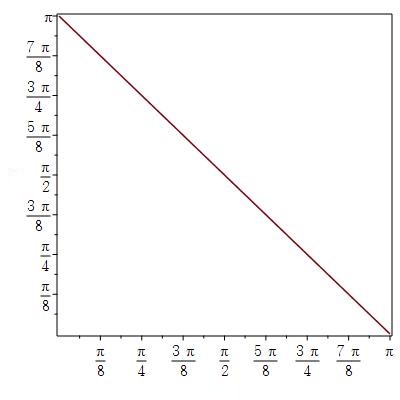}
\hskip 0.2cm
\includegraphics[width=5.2cm]{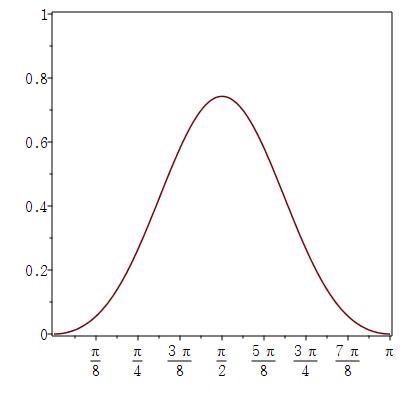}
\hskip 0.2cm
\includegraphics[width=5.2cm]{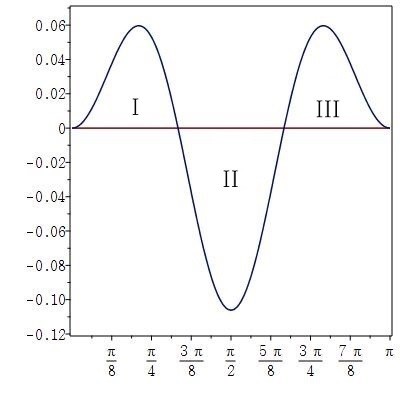}
\caption[chargep] {(a) The profile function $f(\mu)$ in the identity map, and the charge density profiles 
(b) $\rho_{p}(\mu)$ and (c) $\rho_{n}(\mu)$, for proton and neutron respectively, are plotted versus $\mu$. 
Here $f(\mu)$ is related with the BPS bound in the soliton energy, and 
$\rho_{p}(\mu)$ and $\rho_{n}(\mu)$ are associated with the charges inside the nucleons.} 
\label{chargep}
\end{figure}

In this section, we investigate the charge density profiles of nucleons in Fig.~\ref{chargep}(b) and 
Fig.~\ref{chargep}(c). Note that, as in the the profile function in the identity map in Fig.~\ref{chargep}(a), these 
charge density profiles are formulated in terms of the angle $\mu$. 
Moreover the charge density profiles $\rho_{p}(\mu)$ and $\rho_{n}(\mu)$ are associated with the charges inside the nucleons, 
while the profile function $f(\mu)$ in the identity map is related with the BPS bound in the soliton energy.

Now, using the three-sphere metric in (\ref{metrics30}), we construct 
the charge density profiles of proton and neutron
\beq
\rho_{p}(\mu)=\frac{1}{2}\left[\rho_{I=0}(\mu)+ \rho_{I=1}(\mu)\right],~~~
\rho_{n}(\mu)=\frac{1}{2}\left[\rho_{I=0}(\mu)- \rho_{I=1}(\mu)\right],
\eeq
where the isoscalar and isovector electric charge densities are given by~\cite{hong98plb}
\bea
\rho_{I=0}(\mu)&=&-\frac{2}{\pi}\sin^2 f\frac{df}{d\mu}=\frac{2}{\pi}\sin^2 \mu,\nn\\
\rho_{I=1}(\mu)&=&\frac{8\pi}{3{\tilde{\cal I}}}\sin^{2}\mu\sin^{2}f\left(1+\left(\frac{d f}{d\mu}\right)^{2}+\frac{\sin^{2}f}{\sin^{2}\mu}\right)
=\frac{8}{3\pi}\sin^4 \mu,
\label{rhoexp}
\eea
respectively, to produce
\beq
\rho_{p}(\mu)=\frac{1}{\pi}\sin^2\mu \left(1+\frac{4}{3}\sin^2 \mu\right),~~~
\rho_{n}(\mu)=\frac{1}{\pi}\sin^2\mu \left(1-\frac{4}{3}\sin^2 \mu\right).
\label{rhopn}
\eeq
We then find the proton and neutron charges as follows 
\beq
Q_{p}=\int_{0}^{\pi}d\mu~\rho_{p}(\mu),~~~Q_{n}=\int_{0}^{\pi}d\mu~\rho_{n}(\mu),
\eeq 
which yield $Q_{p}=1$ and $Q_{n}=0$ as expected.

The proton and neutron charge densities in  (\ref{rhopn}) are depicted in 
Fig.~\ref{chargep}(b) and Fig.~\ref{chargep}(c), respectively. From  (\ref{rhopn})
we find two root values of $\rho_{n}(\mu)$=0: $\mu=\frac{\pi}{3}$ and $\mu=\frac{2\pi}{3}$. 
We now have three regions I, II and III in the neutron charge density as shown in Fig.~\ref{chargep}(c) and for each region, 
we obtain the charge fractions as follows
\beq
Q_{n}^{\rm I}=\frac{\sqrt{3}}{16\pi},~~~
Q_{n}^{\rm II}=-\frac{\sqrt{3}}{8\pi},~~~
Q_{n}^{\rm III}=\frac{\sqrt{3}}{16\pi},
\label{qn123}
\eeq
which are in good agreement with the result $Q_{n}=0$. Note that the charge density profiles of proton and neutron 
have no dependence on the coordinates $\theta$ and $\phi$ and they have dependence only on the coordinate 
$\mu$ on $S^{1}$.\footnote{Note that for a given fixed value of $\mu=\mu_{0}$, the three-sphere metric in (\ref{metrics3}) becomes
$ds^{2}=\lambda^{2}\sin^{2}\mu_{0}~(d\theta^{2}+\sin^{2}\theta~d\phi^{2})$. This two-sphere metric is given by 
the coordinates ($\theta$, $\phi$) and the fixed two-sphere radius parameter $\lambda\sin\mu_{0}$. 
The charge density profiles of the nucleons thus do not depend on ($\theta$, $\phi$), but depend on $\mu$ for the 
fixed value of $\mu=\mu_{0}$ corresponding to the two-sphere radius parameter $\lambda\sin\mu_{0}$. 
Here and in Sec. IV and Appendix A, we use $\lambda$ instead of $\lambda_{B}$ for the radius parameter of hypersphere, for simplicity.}

\section{Topology and knot structure in HSM}
\setcounter{equation}{0}
\renewcommand{\theequation}{\arabic{section}.\arabic{equation}}

Now we investigate the topological aspects and knot structure in the hypersphere soliton. 
To to this, we first consider the (1+1) dimensional sine-Gordon soliton whose Lagrangian is given in 
appropriate length and energy units
as follows~\cite{skyrme61b,raja,proceeding,manton2}
\beq
L_{SG}=\int dx \left[\frac{1}{2}\pa_{\mu}\phi\pa^{\mu}\phi+\cos\phi-1\right],
\label{sglag}
\eeq
from which we obtain the field equation of the form
\beq
\Box\phi+\sin \phi=0.
\eeq
Note that the sine-Gordon field equation satisfies the discrete symmetry 
$\phi (x,t)\rightarrow \phi (x,t)+2N\pi$, $N=0,\pm 1, \pm 2, \cdots$. 
The topological charge for a static solution of the sine-Gordon kink is then given by 
$Q_{top}=\frac{1}{2\pi}[\phi(+\infty)-\phi(-\infty)]$. We now have the static soliton solutions of the form: 
$\phi (x)=+4\tan^{-1}\exp x$ and $\phi (x)=-4\tan^{-1}\exp x$, for the cases of $Q_{top}=+1$ 
for a kink and $Q_{top}=-1$ for an anti-kink, respectively. These sine-Gordon kink solutions 
represent infinite M\"obius strip feature~\cite{proceeding}. Moreover the sine-Gordon kink state 
is identifiable as a fermion~\cite{raja}. Note in the hypersphere soliton that, exploiting $B$ in  (\ref{bfmu}), we 
have the profile function $f(\mu)=\pi-\mu$ for the hypersphere soliton with the topological charge 
$B=+1$~\cite{hong98plb,manton2}. We also obtain $f(\mu)=-(\pi-\mu)$ for the 
hypersphere anti-soliton with $B=-1$~\cite{manton2}. In this sense, in the hypersphere soliton, we have the M\"obius strip feature similar to that in the sine-Gordon kink.

\begin{figure}
\centering
\vskip -0.5cm 
\includegraphics[width=7.3cm]{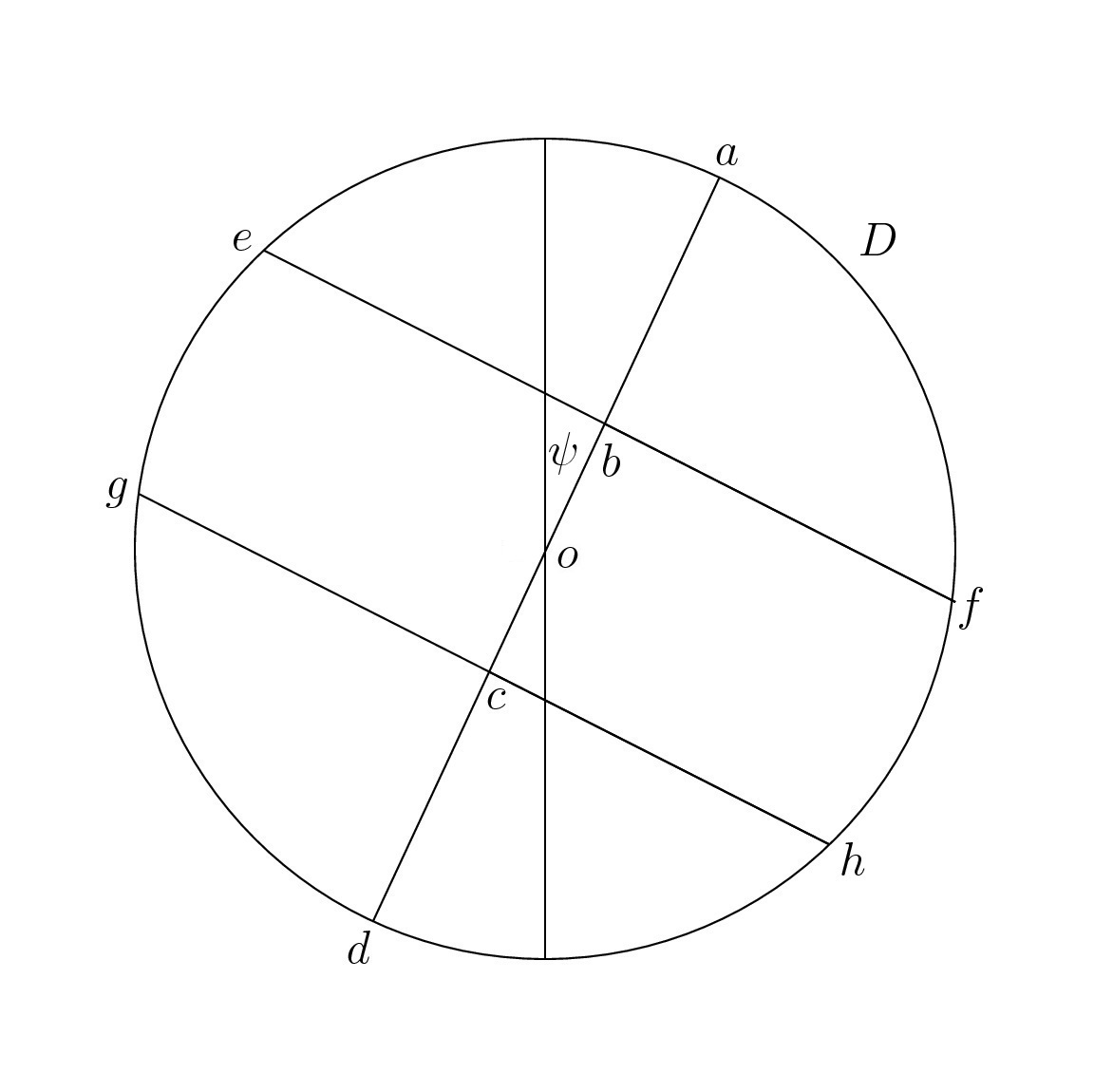}
\vskip -0.5cm \caption[rp2tube] {The cross section of tubular neighborhood of the half-twist circle is delineated by the schematic 
M\"obius strip.}
\label{rp2tube}
\end{figure}

\begin{figure}[h]
\centering
\vskip 0.3cm 
\includegraphics[width=4.0cm]{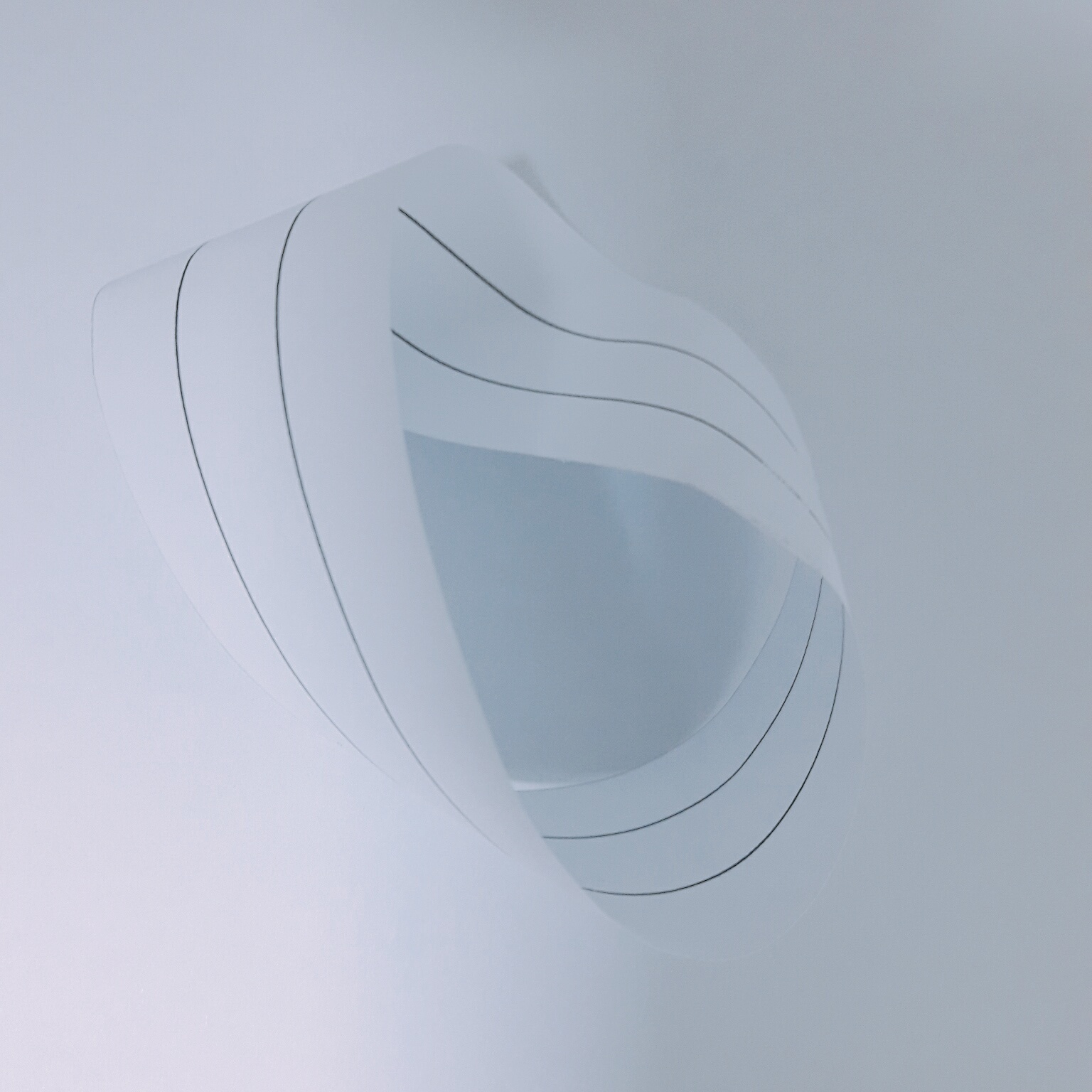}
\hskip 2.0cm
\includegraphics[width=4.0cm]{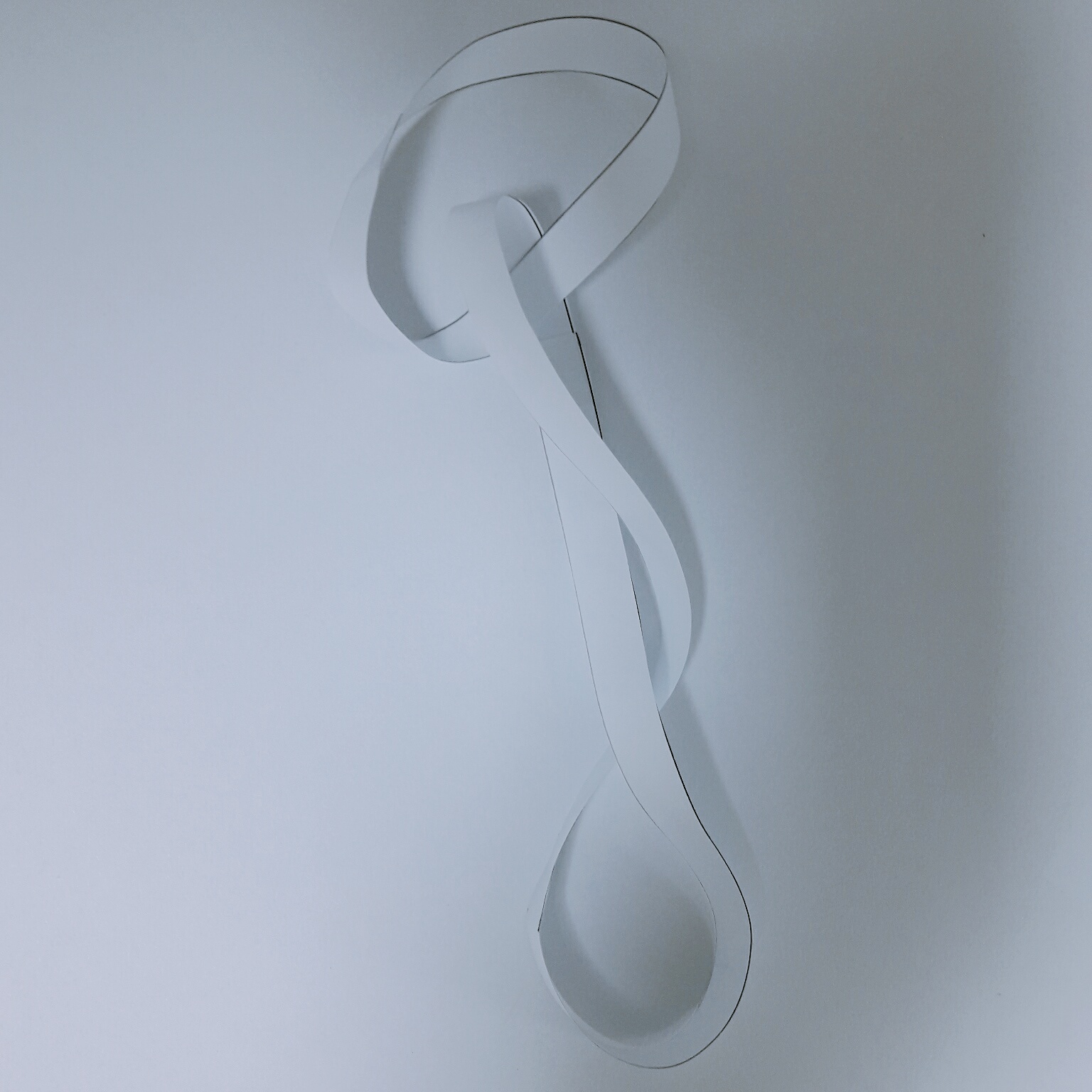}
\vskip 0.0cm \caption[strip] {(a) The schematic M\"obius strip in the HSM corresponds to the fermion, and (b)
the splitted M\"obius strip explains the knot structure of the two different M\"obius strip type circles in the hypersphere soliton.} 
\label{strip}
\end{figure}

Next we consider a {\it half-twist circle} $S^{1}$ inside $S^{3}$ which will be discussed in Appendix A, in order to briefly investigate 
the relations between the three dimensional tubular neighborhood of the half-twist circle and the corresponding M\"obius strip. 
We now assume that the point $o$ is the center of the two dimensional disc $D$ representing the cross section of the tubular neighborhood of the half-twist circle as shown in Fig.~\ref{rp2tube}. Here the line segment $ad$ represents 
the cross section of the M\"obius strip. Next we cut the disc $D$ of the tubular neighborhood into three pieces 
by using the line segments $ef$ and $gh$, and then we let the points $b$ and $c$ denote 
the corresponding cross sectional cutting lines on the M\"obius strip. Here $\psi$ represents the rotation angle of the disc $D$ 
(or the line segment $ad$) with respect to the vertical line which is the initial position of $D$. 
As the disc $D$ rotates along the circle $S^{1}$ of radius parameter $\lambda$, the angle $\psi$ increases from $0$ to $\pi$, 
to complete the cycle of the half-twist. In other words, after the complete rotation the final disc $D$ arrives at the initial disc position, with the angle shift $\pi$. By gluing these two discs together, we obtain a M\"obius strip type tubular neighborhood of the half-twist 
circle $S^{1}$. In particular, the tubular neighborhood is splitted into three pieces with nontrivial volume shapes along the half-twist circle. Moreover the line segment $ad$ inside the tubular neighborhood makes the M\"obius strip. In this respect without loss of generality, 
instead of exploiting the tubular neighborhood of the half-twist circle $S^{1}$ inside $S^{3}$, for visual clarity we will use the 
schematic M\"obius strips in Fig.~\ref{strip}(a) and Fig.~\ref{strip}(b) in the further discussions for the topology and knot structure of the baryons below. We reemphasize that, in Fig.~\ref{strip}(a) and Fig.~\ref{strip}(b), the M\"obius strips represent schematically the tubular neighborhoods of the half-twist circles.

Note that, to describe a fermion such as a nucleon, we need a M\"obius strip with a half-twist, 
associated with a half-twist circle $S^{1}$ inside $S^{3}(=S^{2}\times S^{1})$, as shown in Fig.~\ref{strip}(a). For more details about the 
M\"obius strip related with the fermion, see the last 
paragraph in this section. Here the half-twist circle 
$S^{1}$ plays a major role in the topological structure of the fermion. To relate the HSM to the quark model where, in the fermions 
such as proton and neutron, there exist three quarks, we separate the M\"obius strip into three pieces as shown in Fig.~\ref{strip}(a) to yield two nontrivial M\"obius strips which are Hopf-linked~\cite{kauffman,baez} to each other, as shown in Fig.~\ref{strip}(b). More specifically, we have the M\"obius strip with the circumference length  equal to that of the original 
M\"obius strip in Fig.~\ref{strip}(a) and with a half-twist. This small M\"obius strip originates from the middle piece of the M\"obius strip in Fig.~\ref{strip}(a). Next we have the other M\"obius strip with the circumference length of two times that of the original 
M\"obius strip in Fig.~\ref{strip}(a) and with two twists. This large M\"obius strip consists of the first and third pieces of the M\"obius strip in Fig.~\ref{strip}(a). Note that the two twists of the large M\"obius strip are equal to two times addition of two half-twists 
of the first and third pieces of the original M\"obius strip. Moreover the large strip corresponds to uu in uud and dd in udd for proton and neutron defined in the quark model, respectively. Now we emphasize that two M\"obius strips in Fig.~\ref{strip}(b) are Hopf-linked to produce a knot structure, which will be also discussed in Appendix A. 

Now we return to exploiting the tubular neighborhood of the half-twist circle in the HSM defined on the hypersphere $S^{3}$, instead of using 
the schematic M\"obius strip. As a result, we find that there exists inside the nucleon the half-twist circle $S^{1}(\subset S^{3})$ possessing the M\"obius strip type tubular neighborhood as shown in Fig.~\ref{strip}(a). The tubular neighborhood is now splitted into three pieces along the half-twist circle to yield two Hopf-linked sub-manifolds due to the M\"obius strip characteristic of the tubular neighborhood as shown in Fig.~\ref{strip}(b). Indeed, the two Hopf-linked M\"obius strip type circles in Fig.~\ref{strip}(b) explains the knot structure of the two different M\"obius strip type circles, which correspond to (uu, d) in proton and (dd, u) in neutron, respectively, in the quark model. For more details about the two Hopf-linked circles inside $S^{3}$, see Appendix A. 

Finally we classify the boson and fermion in terms of the strip structures. The boson performs $2\pi$ rotation to return its 
starting situation. The feature is explained by exploiting $2\pi$ rotation on the ordinary strip. 
Next the fermion performs $4\pi$ rotation to return its initial situation according to the relativistic quantum mechanics. 
The characteristic is elucidated by using $4\pi$ rotation on the M\"obius strip shown in 
Fig.~\ref{strip}(a). The baryonic inner structure is thus successfully explained on the hypersphere $S^{3}$. 
In contrast, in the case of bosonic inner structure, we do 
not need to consider such a hypersphere geometry.

\section{Conclusions}
\setcounter{equation}{0}
\renewcommand{\theequation}{\arabic{section}.\arabic{equation}}
\label{conclusion}

In summary, in the HSM we have performed the first class Dirac quantization, to predict more 
rigorously the physical quantities such as the baryon masses and charge radii. To do this, we have exploited the 
identity map associated with the BPS bound in the baryon soliton energy. Moreover we have investigated the charge density profile functions of 
nucleons, and then depicted the density distributions for proton and neutron plotted versus the hypersphere third angle $\mu$. In particular, the neutron charge density has been shown to have the nontrivial $\mu$ dependence, consisting of both the positive and negative charge density parts. Next we have investigated the inner topological structures of the hypersphere soliton, by using the M\"obius strip type tubular 
neighborhood of the half-twist circle in $S^{3}$. In particular, we have shown that, in the HSM, the nucleons are described by the knot of two Hopf-linked M\"obius strip type circles inside the nucleons. As a result, we have found that the 
two Hopf-linked M\"obius strip type circles in the HSM correspond to the (uu, d) in proton and (dd, u) in neutron, respectively, in the quark model.

\acknowledgments{The author would like to thank Professor Peter van Nieuwenhuizen for helpful correspondence and kind encouragement. 
He was supported by Basic Science Research Program through the National Research Foundation of Korea funded by the Ministry of Education, NRF-2019R1I1A1A01058449.}

\appendix
\section{Hypersphere geometry}
\setcounter{equation}{0}
\renewcommand{\theequation}{A.\arabic{equation}}

In order to investigate a hypersphere of radius parameter $\lambda$, we consider the three-sphere metric on $S^{3}$
\beq
ds^{2}=\lambda^{2}d\mu^{2}+\lambda^{2}\sin^{2}\mu~(d\theta^{2}+\sin^{2}\theta~d\phi^{2}).
\label{metrics3}
\eeq
The ranges of the hyperspherical coordinates are given by $0\le\mu\le\pi$, $0\le\theta\le\pi$ and $0\le\phi\le 2\pi$, and 
$\lambda$ ($0\le \lambda <\infty$) is the radius parameter of $S^{3}$. 
On the hypersphere, the three-sphere metric is given by the line element
\beq
d\vec{l}=\hat{e}_{\mu}\lambda~d\mu+\hat{e}_{\theta}\lambda\sin\mu~d\theta+\hat{e}_{\phi}\lambda\sin\mu\sin\theta~d\phi,
\eeq
where $(\hat{e}_{\mu},\hat{e}_{\theta},\hat{e}_{\phi})$ are the unit vectors along the three directions. 

The area elements are given by
\beq
d\vec{a}_{\mu}=\hat{e}_{\mu}\lambda^{2}\sin^{2}\mu\sin\theta~d\theta~d\phi,~~~
d\vec{a}_{\theta}=\hat{e}_{\theta}\lambda^{2}\sin\mu\sin\theta~d\mu~d\phi,~~~
d\vec{a}_{\phi}=\hat{e}_{\phi}\lambda^{2}\sin\mu~d\mu~d\theta,
\eeq
and the three dimensional volume element is defined as
\beq
dV=\lambda^{3}\sin^{2}\mu\sin\theta~d\mu~d\theta~d\phi.
\label{volumnel}
\eeq
In the hyperspherical coordinates system, the gradient operator is given by
\beq
\na=\hat{e}_{\mu}\frac{1}{\lambda}\frac{\pa}{\pa \mu}
+\hat{e}_{\theta}\frac{1}{\lambda\sin\mu}\frac{\pa}{\pa\theta}
+\hat{e}_{\phi}\frac{1}{\lambda\sin\mu\sin\theta}\frac{\pa}{\pa\phi}.
\eeq

Next, we investigate a half-twist circle $S^{1}(\subset S^{3})$ on the hypersphere. 
For the space manifolds $S^{3}$ and $S^{2}$, we have the homotopy 
group~\cite{toda,manton2,schwarz} $\Pi_{3}(S^{2})={\mathbf Z}$. Note that in the homotopy group, the target space $S^{2}$ is a 
geometrical manifold. Due to the above homotopy group, 
we have an associated integer topological charge, namely the 
Hopf charge~\cite{manton2}. Here the Hopf charge is intrinsically different from the topological charge $B$ associated with 
$\Pi_{3}(S^{3})={\mathbf Z}$. In other words, one cannot construct any Hopf charge having a form similar to $B$ in  (\ref{bfmu}). 
Instead, we have the preimage of a point on the target manifold $S^{2}$ which is a closed loop inside $S^{3}$~\cite{manton2}. 
Moreover it is shown in the 
Hopf fibration~\cite{hopf,manton2,bott} that the closed loop is a 
half-twist circle $S^{1}(\subset S^{3})$. To be specific, noting that the homology groups for $S^{3}$ 
and $S^{2}\times S^{1}$ are given by 
$H_{1}(S^{3})=0$ and $H_{1}(S^{2}\times S^{1})={\mathbf Z}$~\cite{derham}, one finds that the hypersphere $S^{3}$ is 
{\it locally} a product of two manifolds $S^{3}=S^{2}\times S^{1}$~\cite{ryder}, implying that the $S^{1}(\subset S^{3})$ is the {\it 
half-twist} circle of radius parameter $\lambda$. It is also interesting to note that the preimages inside $S^{3}$ decompose the hypersphere into a continuous family of circles, and two distinct circles are linked inside $S^{3}$~\cite{manton2,bott} and form the Hopf link~\cite{kauffman,baez}. 


\end{document}